%% file: main.tex
\documentclass[english]{article}

\ifx\pdftexversion\undefined\else
	\pdfoutput=1
	\usepackage[T1]{fontenc}
	\usepackage[utf8]{inputenc}
\fi

\usepackage{microtype}
\usepackage{babel}
\usepackage{lmodern}
\usepackage{csquotes}
\usepackage{mathtools}
\usepackage{amsthm}
\usepackage{amssymb}
\usepackage{hyperref}
\usepackage[style=alphabetic, backref, giveninits]{biblatex}
\usepackage{cleveref}

\numberwithin{equation}{section}
\input{commands}

\begin{document}

\title{The one-legged K-theoretic vertex of fourfolds from 3d gauge theory}
\author{Nicolò Piazzalunga\thanks{NHETC, Rutgers University. Research
supported by the US Department of Energy under grant DE-SC0010008.}}
\maketitle

\begin{abstract}
\noindent
\input{abstract}

\end{abstract}


\input{content}

\printbibliography
\end{document}

%% file: commands.tex
\newcommand{\e}{\epsilon}
\newcommand{\Ga}{\Gamma}
\newcommand{\BZ}{\mathbb Z}
\newcommand{\BR}{\mathbb R}
\newcommand{\BC}{\mathbb C}
\newcommand{\BP}{\mathbb P}
\newcommand{\BS}{\mathbb S}
\newcommand{\Hilb}{\mathrm{Hilb}}
\newcommand{\cont}{\mathrm{cont}}
\newcommand{\cN}{\mathcal N}
\DeclareMathOperator\tr{tr}
\DeclareMathOperator\Hom{Hom}
\DeclareMathOperator\End{End}
\DeclareMathOperator\Res{Res}

%% file: abstract.tex
We present formulas for the K-theoretic Pandharipande-Thomas
vertex of fourfolds, for the case of one non-trivial leg.
They are obtained from computations in a three-dimensional
supersymmetric gauge theory, where we identify the field content
and boundary conditions that correspond to the vertex with
tautological insertions.

%% file: content.tex
\section{Introduction}

Recently, there's been some interest in gauge theories on Calabi-Yau
fourfolds.  These make sense in string theory, as theories on D-branes
when gravity is decoupled, so they're automatically twisted, and are
ultimately justified as non-commutative gauge theories, by taking the
Seiberg-Witten limit.  Generally speaking, the solitons correspond to
D-branes wrapping cycles, and partition functions are Witten indices
counting their bound states.

For toric manifolds, there exists a general machinery due to Nekrasov
that allows to localize such theories and express their partition
functions in terms of simple building blocks, akin to the instanton
partition function of 4d $\cN=2$ theory.  In the present case, this is
replaced by a four-valent K-theoretic vertex.  This is where most of
the complexity of such Donaldson-Thomas counts lies.

Another source of complexity, which is new to four complex dimensions,
is that, although the general machinery is more or less straightforward to
apply, and it produces a decomposition into vertices-edges-faces dictated
by the toric polyhedron, one has to choose signs (orientations) for each
of these, which are not easy to fix.  In other words, the selfduality
equations involve a real representation of $SU(4)$, and one has to take
\emph{half} of the obstruction complex.  This is mirrored in the fact
that one cannot relax the CY condition.

Rather than discussing the general case, here we focus on the one-legged
vertex, where three legs are trivial.  This can be reformulated in terms
of K-theoretic quasimaps counts, as maps from the sphere $\BP^1$ into the
Hilbert scheme of points of $\BC^3$.  While the gauge theory is naturally
DT theory, the quasimap counts are equivalent to Pandharipande-Thomas
theory: this, while being closely related to DT theory, has the advantage
of being more economical, as it factors out contributions from pure
D0-branes.

The corresponding theory for threefolds (not necessarily
CY) has been studied in great generality by the Okounkov
school \cite{Okounkov:2015spn}, building on prior milestones
\cite{Nakajima:1994nid, Nekrasov:2014xaa}.  From there, it is
clear that the object of our study can be looked at from various
angles (with some subtle differences): GLSM partition functions
on the disk and D-brane central charges \cite{Hori:2013ika},
holomorphic blocks \cite{Beem:2012mb}, and Givental's I-function
\cite{givental1996equivariant}.

Recently, the mathematical foundations of quasimap theory relevant
to DT4 have been developed \cite{Cao:2023lon}, building on previous
works \cite{Ciocan_Fontanine_2010, park2021virtual}.  Physically, the
K-theoretic PT vertex with one non-trivial leg, with plane partitions
classifying fixed points on $\Hilb^k(\BC^3)$, and with tautological
insertions, should be equal \cite{Kapustin:2013hpk} to the partition
function of a topologically twisted 3d $\cN=2$ theory on the cigar.
Here, we argue that one can use the physical $\cN=2$ theory on $D \times
\BS^1$, with $D$ being the disk, to compute certain observables of the
topologically twisted theory, upon making the correct identifications
\cite{Jockers:2018sfl}.  In the physical theory, it's clear that
tautological insertions, corresponding to fundamental matter in DT theory,
are chiral fields with Dirichlet boundary conditions.  The point is that
one needs to cancel anomalies for Chern-Simons terms, and one way to
do it is via anomaly inflow from the boundary \cite{Yoshida:2014ssa}.
This is crucial to obtain the correct measure for the PT vertex, which
is our main result in \cref{3dpt}.

Despite their physical origin, our formulas are mathematically rigorous,
and the integration contour is completely determined, so we do not need
to make any ansatz for it.

\section{The setup}

Let us consider stable quasimaps from $C=\BP^1$ to $\Hilb^k (\BC^3)$,
twisted by the line bundles $L_a$ over $C$ of degree $l_a$, such that
$K_C = L_1 \otimes L_2 \otimes L_3$, and $l_1+l_2+l_3 = -2$.  Denote by
$N$ the (fixed) vector bundle of rank $n=1$ over $C$.  Let $G=U(k)$.
A quasimap is the datum of a principal $G$-bundle over $C$ and a section
$f= (B_1,B_2,B_3,I)$ of the associated rank-$k$ complex vector bundle $K$,
where $B_a \in H^0(C,\End(K) \otimes L_a)$ and $I \in H^0(C,\Hom(N,K))$.
The degree is $\deg f = c_1 (K)$.

The corresponding topological field theory is obtained by twisting the
$\cN=(2,2)$ theory on $C$.  The twist identifies the surviving R-symmetry
with the $U(1)$ Lorentz isometry on $C$.  The fields $X=(B_1,B_2,B_3,I)$
of the 2d cohomological theory satisfy equations \begin{equation} D_{\bar
z} X = -G^{-1} \frac{\partial \bar W}{\partial \bar X} \end{equation}
which follow from the superpotential $W = \tr B_1 [B_2,B_3]$.  The fields
$B_a$ have R-charges $-l_a$, so $W$ is quasi-homogeneous of degree two.
Here $z$ is a local coordinate on $C$, and $G$ a Kähler metric.  We can
either quotient by $GL(k)$, or quotient by $U(k)$ if we replace the
stability condition by the D-term equation \begin{equation} * F + \mu
(X,\bar X) = \xi \end{equation} where $F$ is the field strength on $C$,
and $\mu$ the momentum map associated to the symplectic form determined
by $G$.  The K-theoretic counts correspond to the three-dimensional lift
of the 2d twisted theory.

We consider quasimaps with a non-singular boundary condition at $\infty
\in \BP^1$, and compute the K-theoretic push-forward of tautological
insertions using the evaluation map at $\infty$ and the orientation
defined by the virtual fundamental class, and weighted by $z^{\deg f}$.
We work with the physical three-dimensional $\cN=2$ theory on $D \times
\BS^1$, with gauge group $U(k)$ and the following field content: 3d
vector, 3d chirals with Neumann boundary conditions ($B_a$, $I$), 3d
chiral with Dirichlet boundary conditions (corresponding to fundamental
matter, i.e.\ tautological insertions in the quasimap theory), zero bare
Chern-Simons level, and (crucially) 2d vector to cancel the anomaly.
The $\Omega$-background acts on $\BR^2 \times \BS^1$ as $q=e^{\beta \e}$,
where $\beta$ is the radius of the circle $\BS^1$, and it's natural to
identify this parameter with the fugacity for the sum of $U(1)$
R-charge and rotations generators, denoted by $q^2$
in ref.~\cite{Yoshida:2014ssa}.

\section{The computation}

We work equivariantly with respect to all symmetries.  The torus
$(\BC^*)^3$ acts on $\Hilb^k (\BC^3)$, with fixed points in one-to-one
correspondence with plane partitions of size $k$.  We turn on twisted
masses $q_a=e^{\beta\e_a}$ for $B_a$.  For the disk, we do not need
to worry about R-charges (while they matter for the sphere case),
as we can reabsorb them in the equivariant parameters. We must impose
the CY4 condition $q \prod_{a=1}^3 q_a = 1$.  The equivariant parameter
$a=e^{\beta \alpha}$ acts on the framing node $N$, while $\mu=e^{\beta m}$
is a mass for the fundamental matter.

For $|q|<1$, the K-theoretic disk partition is computed by the residue
\begin{equation} Z_k = \frac1{k!} \oint_{|x_i|=1} \phi (-T_k) \,
\prod_{i \neq j} \theta (x_i/x_j) \prod_{i=1}^k \frac{dx_i}{2\pi
i x_i} x_i^{-\xi} \end{equation} where the functions $\phi$ and
$\theta$ are defined in \cref{formulas}, and \begin{equation} T_k =
\sum_i x_i (a^{-1} - {\mu}^{-1}) + \sum_{i\neq j} \frac{x_i}{x_j}
(q_1^{-1}+q_2^{-1}+q_3^{-1}-1) \end{equation} is determined by the
field content.  We require that $|a|,|q_1|,|q_2|,|q_3|>1$, so that the
denominator is convergent.

\paragraph{Contour}  For $\xi > 0$, the contour picks up poles at one of
the hyperplanes \begin{equation} x_{i} = a q^{-d_1}, \quad
d_1 \geq 0 \end{equation} and $k-1$ of the hyperplanes
\begin{equation} \frac{x_r}{x_s} q_a^{-1} = q^{-d_p}, \quad d_p \geq 0,\,
a=1,2,3,\, p=2,\ldots,k \end{equation} such that $x=x_{cl}$,
defined by taking all $d=0$, is the character of a plane partition in
$q_1$, $q_2$, $q_3$.  (We call any such $x_{cl}$ a classical pole.)

For reference, the disk partition function \cite{Hori:2013ika} reads
\begin{equation} F_k = \frac1{k!} \oint
\prod_{i=1}^k \frac{\Ga(\phi_i-\alpha)} {\Ga(\phi_i-m)} \prod_{i \neq
j} \frac{ \prod_{a=1}^3 \Ga(\phi_i-\phi_j-\e_a)}{\Ga(\phi_i-\phi_j)}
\prod_{i=1}^k \frac{d\phi_i}{2\pi i} \, e^{-\phi_i \xi} \end{equation}

\section{The conjecture}

Let us define the content of plane partition $\pi$ corresponding to
pole $x_{cl}$ \begin{equation} \cont (\pi) = \prod_{i=1}^k x_{cl,i} =
\prod_{(i,j,k)\in \pi} a q_1^{i-1}q_2^{j-1} q_3^{k-1} \end{equation}
and the perturbative prefactor \begin{equation} Z_{pert} (\pi) =
\cont(\pi)^{-\xi} \, \phi(k - T_k(x_{cl}) - qk) \, \prod_{i \neq j}
\theta (x_{cl,i}/x_{cl,j}) \end{equation}

Upon identifying $z=q^{\xi}$ as the box-counting parameter, we claim
that \begin{equation} \label{3dpt} Z_k = \sum_{|\pi|=k} Z_{pert} (\pi)
\, V_{PT} (0,0,0,\pi) \end{equation} where $V_{PT}$ is the vertex of
PT theory and the sum runs over plane partitions.
We checked this claim against conjectural formulas for the
general four-vertex of Donaldson-Thomas theory in the upcoming work with
N.Nekrasov, for the cases $|\pi| \leq 2$ and the first few instantons.

\section{Conclusion and outlook}

We presented formulas for the K-theoretic PT vertex in the case of one
non-trivial leg, constructed from the supersymmetric gauge theory in three
dimensions.  We exploited the physical theory to make sense of fundamental
matter in terms of boundary conditions, and used the physically-motivated
anomaly cancellation to get the correct vertex measure.  We recovered
the $\Omega$-background of the twisted theory from the fugacity for the
appropriate combination of R-charge and Lorentz rotation.

Although we focused on the important example of $\Hilb (\BC^3)$, it
is possible to extend this procedure to other quivers with potential,
by either computing the virtual tangent space or equivalently the
corresponding field content.

One topic that deserves further investigations are the difference
equations satisfied by the vertex.  The first type comes from the
operation of shifting the Kähler parameter, $\xi \to \xi+1$, which
corresponds to inserting the observable $\prod_i x_i \sim \det K$.
They work in the same fashion as the abelian case.  The second set is
more interesting, and it comes from acting on the equivariant parameters,
as $q_a \to q_a q^{-l_a}$, $q \to q^{-1}$, from the degeneration idea.
The $R$-matrix for $\Hilb^k(\BC^3)$ is a $p_k \times p_k$ matrix, where
$p_k$ is the number of plane partitions of size $k$, but this is not
enlightening, unless we know that it can be reduced to smaller blocks.

\paragraph{Acknowledgments} We thank D.-E.Diaconescu, N.Nekrasov and
M.Romo for inspiring discussions, as well as W.-y.Chuang, M.Porta and
F.Sala for sharing their knowledge on the mathematical foundations.
We thank A.Grassi for playing with us on the $k=1$ case, where we've
been stuck for some time.

\appendix

\section{Definitions and elementary properties} \label{formulas}

Let $(a;q)_n = \prod_{k=0}^{n-1} (1-a q^k)$, with $(a;q)_0=1$.
Let $\phi(a) = (a;q)_\infty$, and extend it on K-theory classes by
multiplication.  It satisfies the difference equation $\phi(qx) =
\frac{1}{1-x} \phi(x)$.


We extend the definition to all integers by the relation \begin{equation}
(x;q)_p := \frac {\phi(x)} {\phi(xq^p)} =\begin{dcases*} (x;q)_p &
if $p\geq 0$ \\ \frac{1}{(xq^p;q)}_{-p} & if $p <0$ \end{dcases*}
\end{equation}

Let us denote $\pi_d  = \prod_{i=1}^d (1-q^{-i})$ for $d \geq 0$ and
$(x)_p = (x;q)_p$ for $p \in \BZ$.

Let us define $\theta(y) := \theta(y;q) = \phi(y) \phi(y^{-1}q)$.
This satisfies the relation \begin{equation} \theta(yq^p) = \theta(y)
(-1)^p y^{-p} q^{-p(p-1)/2} \end{equation} for $p \in \BZ$, as well as
the relation $\theta(y^{-1}) = -y^{-1} \theta(y)$.

The residue evaluation, for $k \in \BZ_{\geq 0}$, gives \begin{equation}
\Res_{y=q^{-k}} \frac{dy}{2\pi i y} \phi(-y) =(-1)\frac {\phi(-q)}
{\prod_{i=1}^k (1-q^{-i})} =(-1)^{k+1} \frac {\phi(-q) q^{k(k+1)/2}}
{(q;q)_k} \end{equation}